%% file: paper-2col.tex
\documentclass[sigconf, nonacm, 10pt]{acmart}
\usepackage[english]{babel}
\usepackage{blindtext}
\usepackage{algorithm}
\usepackage{algorithmic}
\usepackage{makecell}
\usepackage{float}
\usepackage{subfigure}
\usepackage{graphicx}

\renewcommand\footnotetextcopyrightpermission[1]{} 
\acmPrice{}
\acmConference[CoRR]{}
\acmYear{2021}
\copyrightyear{}

\begin{document}
\title{Third-party Service Dependencies and Centralization Around the World}

\author{Rashna Kumar \quad Sana Asif \quad Elise Lee \quad Fabi\'an E. Bustamante}
\affiliation{Northwestern University}
\renewcommand{\shortauthors}{R. Kumar et al.}

\begin{abstract}
There is a growing concern about consolidation trends in Internet services, with, for instance, a large fraction of popular websites depending on a handful of third-party service providers. In this paper, we report on a large-scale study of third-party dependencies around the world, using vantage points from 50 countries, from all inhabited continents, and regional top-500 popular websites. This broad perspective shows that dependencies vary widely around the world. We find that between 15\% and as much as 80\% of websites, across all countries, depend on a DNS, CDN or CA third-party provider. Sites’ critical dependencies, while lower, are equally spread ranging from 9\% and 61\% (CDN and DNS in China, respectively). Despite this high variability, our results suggest a highly concentrated market of third-party providers: three third-party providers across all countries serve an average of 91.2\% and Google, by itself, serves an average of 72\% of the surveyed websites. We explore various factors that may help explain the differences and similarities in degrees of third-party dependency across countries, including economic conditions, Internet development, language, and economic trading partners.

\end{abstract}

\maketitle 

\input{sections/introduction}

\input{sections/motivation}

\input{sections/methodology}

\input{sections/results}

\input{sections/trendAnalysis}
\input{sections/discussion}

\input{sections/relatedwork}

\section{Conclusions and Future Directions}

We presented the first large-scale study of third-party dependency and centralization around the world. We used vantage points in 50 countries, across all inhabited continents, and focused on regional top-500 most popular websites. The main observations is that, while third-party service dependencies seems comparatively high in developed, English-speaking economies, with high Internet development, there is a wide range of third-party service dependencies ($>=$35\%) across all countries and, perhaps most problematic, high levels of centralization across the board. Third-party reliance across countries varies from as low as 14.8\% for CDNs to as high as 71.8\% in DNS and 79.6\% in CAs. Despite this variability in dependency, we find high degree of centralization across all services and countries surveyed. Centralization on top-3 DNS, CA or CDN providers is above $\approx58\%$ (57.9\% on DNS in Hungary) and as high as 87.7\% (DNS in Taiwan) and 95.3\% (CDN in China). 


There are a number of promising directions for future work, some of which we have begun to explore. We plan to expand our vantage points to gain better coverage of certain regions in the world, especially the Middle East, Africa and Latin America.  We have shown that there is value in a country-level analysis of Internet infrastructure dependencies, which can identify trends that a broader global analysis would miss. We are currently automating our analysis tool with the goal of creating periodic checkpoints of these trends around the world. 

\bibliographystyle{ACM-Reference-Format}
\bibliography{base,reference}
\input{sections/appendix}

\end{document}

%% file: sections/introduction.tex
\section{Introduction}
\label{sec:introduction}

There is a growing concern about consolidation trends in the Internet -- the concentration of traffic, infrastructure, services and users on a handful of providers - given its economic, political and reliability implications~\cite{arkko:centrality,huston:dns-centrality,huston:cdn-centrality,ietf:consolidation}. Recent work has started to analyze these trends in the context of DNS~\cite{zembruzki2020measuring,allman:imc18}, cloud providers~\cite{moura2020clouding} and the Web in general~\cite{kashaf2020analyzing}.

Centralization could amplify the impact of vulnerabilities and faults from the shared services, increase the risk of a captive market and significantly reduce user privacy by exposing a more complete user profile to a few, cross-market providers. The Dyn DDoS attack of 2016, the GlobalSign Certificate Revocation Error of 2016, the Amazon DNS DDoS attack in 2019, and the recent 2021 outage of Facebook and subsidiary companies, are just a handful of illustrating examples~\cite{globalsign,dynattack,cloudflare:down,azuredns,godaddyca,cloudflarecdn,facebookoutage}.

The Web provides a relatively accessible environment to characterize these centralization trends and, incidentally, to understand their concerning implications. Access to a website depends on a number of services provided by third-parties such as DNS, Content Distribution Networks (CDNs), and Certificate Authorities (CAs). Accessing a site requires contacting at least one DNS authoritative nameserver to find the IP address of the webserver (or servers) hosting the site's content. These servers could be run by one or more CDNs, for increased reliability and performance. If these hosting servers use HTTPS, a client may also need to contact one or more CAs to verify the validity of the servers' SSL certificates. A popular website  could rely on third-party providers for all these critical services. 

A recent study by Kashaf et al.~\cite{kashaf2020analyzing} leverages this observation to evaluate third-party dependencies in popular websites. The study characterizes the dependency of Alexa's most popular sites on third-party DNS, CA and CDNs, based on data collected from a single vantage point in the US. The authors' analysis reveals, among other concerning findings, that between 78-97\% of Alexa's top-100k websites depend on third-party DNS, CA or CDN, and that the use of third-party services is highly concentrated with the top-3 CDN, DNS or CA providers serving 50-70\% of all websites they surveyed. 

We build on this and other prior work \textit{\textbf{to explore if, and to what extent, third-party dependency and centralization varies across countries and regions of the world}}. 

Our work is motivated by two simple observations (\S\ref{sec:Motivation}). First, while websites could potentially be accessed anywhere, not all websites are popular everywhere. Indeed, the popularity of websites, beyond a few top-ranked ones, is known to be region specific. 


Second, while many third-party services such as DNS and CDNs have been building global infrastructures, on and off-networks~\cite{gigis:hypergiants}, their deployment is not (yet) omnipresent and their relative performance with respect to their competition varies across markets. This has served as motivation for brokering CDN~\cite{biliris:cdnbrokering} and, more recently, multi-cloud architectures~\cite{yeganeh:multicloud,singh:multicdn18}. 

Motivated by these observations and concerns, our work investigates third-party and centralization trends in the wider Internet. We report on a large-scale study of third-party dependencies around the world, using vantage points from 50 countries, across all inhabited continents, and regional top-500 popular websites for a total of 15,774 unique sites. 

We find that between 15\% and as much as 80\% of websites, across all countries, depend on a DNS, CDN or CA third-party provider. Although lower, critical dependencies -- cases where the availability of a service affects the accessibility of a website -- are equally spread ranging from 9\% (CDN in China) to 61\% (DNS in China). The market of third-party providers, however, seems highly concentrated: the top-three third-party providers across all countries serve an average of 91.2\% of all sites (minimum of 56.2\%) and Google, \textit{by itself}, serves an average of 72\% of all websites. We explore various factors that may help explain the differences and similarities in levels of third-party dependency across countries and regions, including economic conditions, Internet development, primary official language, and economic trading partners. 

In summary, we make the following contributions:
\begin{itemize} 
	\item We describe a methodology to carry on a large-scale analysis of third-party dependencies and centralization around the world (\S\ref{sec:Methodology}).
	\item We apply this methodology to characterize the dependencies of top regional sites in 50 countries, together capturing $\approx78\%$ of the Internet user population and different levels of uniqueness of their top regional sites. We find that third-party dependencies and critical dependency around the world vary widely, but market concentration is surprisingly high across all countries (\S\ref{sec:Findings}).
	\item We explore various factors that may help explain the observed differences, from economic conditions to language and trading partners (\S\ref{sec:trendAnalysis}).
	\item We make available the collected dataset, including the list of top sites, and analysis code for public use to enable replicability.
\end{itemize}

%% file: sections/motivation.tex
\section{Background and Motivation}
\label{sec:Motivation}

Our work is motivated by recent reports about consolidation trends in the Internet~\cite{zembruzki2020measuring,allman:imc18,moura2020clouding,kashaf2020analyzing,gigis:hypergiants} and their economic, political and reliability implications~\cite{arkko:centrality,huston:dns-centrality,huston:cdn-centrality,livingood2019centralized,zembruzki2020measuring,ietf:consolidation,arkko:cyberpolicy}. The 2019 Global Internet Report~\cite{ietf:consolidation} provides an overview of these trends in every aspect of the Internet economy, from access provision to service infrastructure to applications. It points out that, while consolidation is often seen as an expected result of maturing markets and industries, the combination of society's increased dependency on the Internet, business agility, and the almost total lack of regulation, is leading to a handful of platforms (sometimes referred to as ``GAFAs'' -- Google, Apple, Facebook, and Amazon, and ``BATs''-- Baidu, Alibaba, and Tencent) in control of much of the Internet's functionality and interoperability. 

The Web offers an accessible context where to characterize and potentially monitor these trends, building on third-party platforms for all its key services, from DNS and CDNs, to advertisements, user tracking and CAs. A client visiting a website needs to interact with each of these services -- to find the IP of servers hosting the content, validate the servers' credentials, and fetch the content many times from different CDNs. Kashaf et al.~\cite{kashaf2020analyzing} leverages this observation to study the level of dependency of popular websites on third-party services, and the consolidation of these markets. Their work reveals several interesting trends. For instance, between 78-97\% of Alexa's top-100k websites depend on the third-party DNS, CA or CDN, and third-party dependencies are higher for less popular sites. Critical dependencies is also high with 84.8\% of the top 100k sites critically depending on a third-party DNS, a dependency that increased by 4.7\% between 2016 and 2020. In terms of consolidation trends, the study reveals a highly concentrated market with 4 out of 10k DNS providers and 2 out of 86 CDN providers critically serving 50\% of websites.

\begin{figure}[ht]
	\centering
	\includegraphics[width=\linewidth]{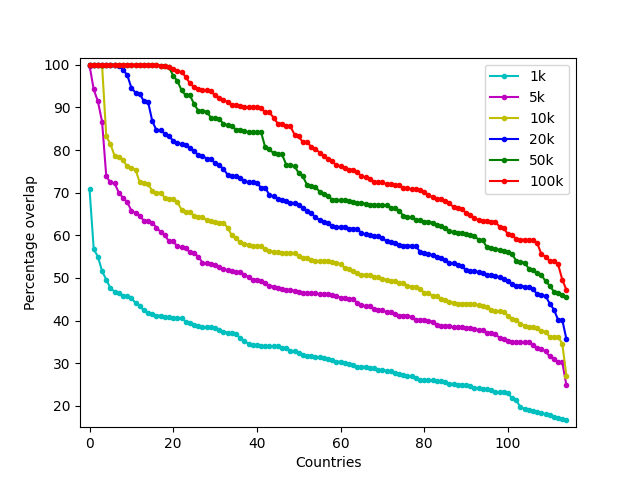}
	\caption{Overlap between Alexa global sites and top regional sites. Countries are sorted from high to low based on their overlap and plotted left to right. $\approx$50\% of countries have an overlap $<77\%$ between their top-500 regional and the top-100k global sites.}
	\label{fig:overlaps}
\end{figure}

Our work is motivated by two main observations. First, the Web is made of 1.9 billion websites as of October 2021~\footnote{https://www.internetlivestats.com/total-number-of-websites/} and except for a few global websites, most sites are not universally popular. Second, while the infrastructure of large third-party services continues to expand, it is not yet omnipresent and may fare differently against their competition at different locales. To illustrate the first point, we collected the top-500 regional sites for each of the 115 countries from Alexa\footnote{Despite concerns with the representativeness and stability of Alexa rankings, these are the longest lists of regional website rankings and considered the appropriate choice for end-user-based analysis~\cite{oliver:toplists}.}  and the top 100k sites from the overall Alexa ranking. For each of these countries, we measured the overlap of the top-500 regional sites with the top 1k, 5k, 10k, 20k, 50k and 100k subsets of the top global sites. As we extend the list of global sites considered, there is a higher probability of including a regional popular site as part of the global ranking and, thus, of finding a higher overlap between global and regional lists. 

Figure~\ref{fig:overlaps} shows the percentage of overlap between these sets. The overall top-ranking websites are clearly dominated by just a few countries, even with the largest global ranking subset, with $\approx~$25\% of them having an overlap higher than 94\%. On the other hand, half of the countries have less than $\approx$77\% overlap even with the full top-100k list of websites. The second observation has served as motivation for brokering CDN~\cite{biliris:cdnbrokering} and multi-cloud architectures~\cite{yeganeh:multicloud,singh:multicdn18}. As no single infrastructure is omnipresent, websites popular in specific regions may choose to build on their own services or contract a well-provisioned local service provider. Even globally popular content providers (CP) are known to contract with different CDNs and DNSs in different part of the world based on connectivity, availability or cost~\cite{singh:multicdn18}. For instance Rakuten, the Japanese electronic commerce and online retailing company, uses Akamai as its DNS provider for rakuten.com, but a private DNS for rakuten.jp. Similarly, wikipedia.org uses DigiCert as its CA provider in Germany and Sweden but IdenTrust Inc. in the US and Canada and microsoft.com hosts content on both Akamai and Amazon Cloudfront in Canada but only on Akamai in Sweden.
 
Combined, these observations motivate the need to investigate if, and to what extent, third-party dependency and centralization varies across countries and regions of the world, and argue for a measurement approach that $(1)$  focuses on regional top-ranked websites, and $(2)$ relies on vantage points around the world. We describe our data collection and analysis methodology in the following sections.

%% file: sections/methodology.tex
\section{Measurement Methodology and Dataset}
\label{sec:Methodology}

In this section, we describe the measurement methodology we employ to characterize service dependencies and centralization around the world. We first select a set of countries for the analysis based on the degree of overlap between the Alexa's top-100k global ranking and the country's top-500 websites\footnote{The maximum number of websites included in regional rankings.}. To properly characterize the service dependency of websites in a given country, we must access these websites as would a user located in that country, as the actual services and resources served through a site may depend on the user's location. To this end, we identify a set of VPN vantage points from a selected set of countries, taking several steps to gain confidence in their reported location. We launch measurements from each of these VPN vantage points to their country's set of top-500 regional websites, focusing on three major services -- DNS, CA and CDN. For each of these services we rely on several heuristics to label each as an internal, third-party or critical service. We consider a service to be critical for a specific website if the service is provided by a third party and the unavailablity of the provider can cause the service being denied.

We begin by describing the approach for selecting the countries included in this study.

\subsection{Country Selection}

For our analysis, we select a subset of all possible countries ensuring: $(1)$ that the overlap between the top-500 regional sites for the included countries and the global-ranked lists  covers the range of possible overlaps, $(2)$ that we can get access to VPN vantage points in those countries and that the claimed location of such vantage points can be verified, and $(3)$ that, together, the sample of countries in the analysis capture a sufficiently large fraction of the Internet user population. 

We first classify all 115 countries with regional rankings in Alexa based on the degree of overlap between their regional sites and the list of top global sites. We divide the set of countries using this overlap into three groups: \textit{high-overlap}, defined as the top third of countries with the highest overlap, \textit{medium-}, the medium third and \textit{low-overlap}, the bottom third, and select countries from each group to ensure different degrees of \textit{representation} in the global ranking. Of the select 50, listed in Table~\ref{tab:overlap} with their two-letter abbreviations, 38.4\%, 25.9\% and 35.7\% have a high, medium and low-overlap respectively.

\begin{table}[ht]
	\centering
	\caption{Countries with different degrees of overlap between their Top-500 Alexa regional sites and the global ranking (The mapping of country Code to their Country Names is included in the appendix).} 
	\label{tab:overlap}
	\begin{tabular}[tp]{ |l|l| }
		\hline
		Overlap Class & Country Codes\\
		\hline \hline
		High & \makecell[l]{AE, AR, AU, BR, CA, CN,  DE, ES, \\
			FR, GB, GR,  HK, ID, IN, IT, JP, \\
			KR, MX, MY, SG, TR, TW, US, VN} \\
		\hline
		Medium & \makecell[l]{BE, CH, CL, CR, IL, UA, PL, NL,\\ 
			NO, RO, SE, TH, ZA} \\
		\hline
		Low & \makecell[l]{AL, BA, BG, CZ, DK, EE, GE, HU, \\
			LV, MD, NZ, PT, RS}\\
		\hline
	\end{tabular}
\end{table}

\begin{figure}[ht]
\centering
\includegraphics[width=\linewidth]{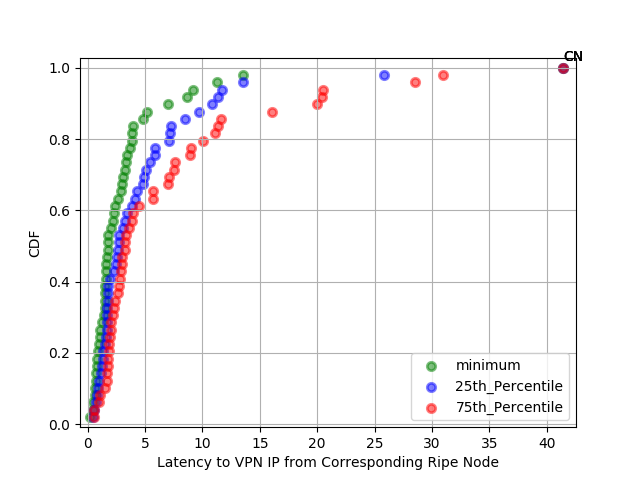}
\captionof{figure}{Latency (ms) from Ripe Atlas Nodes to corresponding VPN Nodes.}
\label{fig:VPNLatency}
\end{figure}

Of the 50 countries, all but China have vantage points available through the Nord VPN. To include China we use the Hotspot Shield VPN, since Nord and many other VPNs do not provide service in China~\cite{narseo:vpnecosystem}. To gain confidence on the claimed location of the vantage points, we obtain the vantage points' public IPs (resolving a domain whose authoritative server we control) and use a set of five Ripe Atlas nodes in the same country to send a sequence of three ICMP pings to the vantage point. Figure~\ref{fig:VPNLatency} shows the minimum, 25th and 75th percentile of ping latency from the set of Ripe Atlas nodes to the corresponding vantage point. The minimum latency to all vantage points, with the exception of the one in China, is within 15ms and 86\% of minimum latencies are below 5ms, suggesting that these nodes are within the claimed country. We further geolocate the vantage points' IPs using two popular geolocation databases: MaxMind GeoLite2~\cite{maxmind} and IP2Location BD11.Lite~\cite{ip2location}. Past work has shown the geolocation databases to be reliable at the country-level~\cite{poese:ipgeolocationdb}. Both databases place the IPs of all 49 Nord VPN nodes in the claimed countries. China proved to be more challenging. The ping latency to the VPN node in China are relatively high at 41ms, although still reasonable for a country of the size of China. The two geolocation databases we employed also disagree with Maxmind geolocating the node in China, but IP2Location database placing it in Japan. 


Last, we group the selected countries in six different regions: America, Europe, Asia Pacific, Africa, and the Middle East and calculate the percentage of the world's Internet population covered by the corresponding countries in those regions. Table~\ref{tab:internetusers} lists the different regions, number of countries included in our dataset, and the percentage of Internet covered. In total, our dataset covers all inhabited regions of the world and captures 78.1\% of the world's Internet users.

A table with the complete set of 50 selected countries, their two-letter codes and percentage of the world's Internet population is included in the Appendix.



\begin{table}[ht]
\centering
\caption{Summary of coverage per region -- number of countries percentage of the world's Internet population.} 
\begin{tabular}[tp]{ |l|c|c| }
	\hline
	Region & Countries & \% of World\\
	\hline \hline
	America & 7 & 13.9\\
	\hline
	\makecell[l]{Northern Europe} & 6 & 1.9\\
	\hline
	\makecell[l]{Western Europe} & 5 & 3.7\\
	\hline
	\makecell[l]{Eastern Europe} & 8 & 2.2\\
	\hline
	\makecell[l]{Southern Europe} & 7 & 2.6\\
	\hline
	\makecell[l]{Asia Pacific} & 13 & 51.4\\
	\hline
	\makecell[l]{Africa, and \\the Middle East} & 4 & 2.4\\
	\hline
	Total & 50 & 78.1\\
	\hline
\end{tabular}
\end{table}
\label{tab:internetusers}

\subsection{Data Collection Methodology}

Using VPN vantage points in the set of selected countries, we launch measurements from these vantage points to their country's set of top-500 regional websites, and use a range of heuristics for labeling three major services -- DNS, CA and CDN.
 
\subsubsection{DNS Measurements.}

We first use a number of heuristics to label all nameservers used by a website as \textit{unknown}, \textit{private} or \textit{third-party}. For each website, we found all the nameservers used by the website by issuing a NS query to the domain name. For each nameserver used by a website, we begin by labeling that nameserver as of an \textit{unknown} type. We then compare the top level domain (TLD) of the website and the nameserver; a match of these two would indicate a \textit{private} nameserver~\cite{kumar:www17}. While the \textit{TLD-matching} heuristic works well in most cases, it may result in misclassifications of some nameservers. For instance the nameserver of youtube.com is *google.com and though both are served by the same provider, the \textit{TLD-matching} heuristic will classify the nameserver as third-party. 

To resolve the above issue we make use of a additional heuristics based on Subject Alternative Names (\textit{SANs List})~\cite{butkiewicz:imc11}. If the website uses HTTPs, we find the site's SANs list via the SSL certificate of the website. For each unclassified nameserver, we then look for the presence of the top-level domain of the nameserver in the SAN list, whereby the presence indicates a \textit{private} DNS nameserver. This heuristic correctly identifies youtube.com using a private DNS.

We use a third heuristic based on Source of Authority records (SOA) -- \textit{SOA-record-matching} -- to label the unclassified nameserver~\cite{kashaf2020analyzing}. In this case, we compare the entity pointed to by the SOA records of the website and the DNS provider pointed to by the SOA records of the nameserver; a mismatch here indicates \textit{third-party} DNS nameserver. 

For the remaining unknown nameservers, if the concentration of the nameserver is larger than 50\footnote{Following Kashaf et al.~\cite{kashaf2020analyzing}.}, we label it as \textit{third-party}.

The second condition of Algorithm \ref{alg:code2} (\S\ref{sec:appendix}) summarizes our heuristic when the service type is instantiated as \textit{DNS}.  This basic three step classification logic, involving TLD-matching, SANs-List and SOA-record-matching, is described in Algorithm \ref{alg:code1} (\S\ref{sec:appendix})  where the \textit{service.url} is provided to the algorithm is the DNS nameserver.

\paragraph{DNS Redundancy} We additionally measure websites that are redundantly provisioned by measuring the percentage of websites that use multiple unique DNS providers across each country. We also measure the percentage of websites that are served by a single DNS provider (i.e. critically dependent on that DNS provider), served by multiple third-party DNS providers and the percentage of websites that are served by private DNS providers and third-party DNS providers.


\subsubsection{CDN Measurements.}
\label{subsec:cdn}

To find set set of CDNs hosting the targeted website and determine whether the CDNs used are private or a third-party service, we find the CNAME of the internal resources of the website. We first find all the internal resources of the website. We use the webdriver capabilities of the Selenium library in python to generate a HAR file for each website which gives us all the resources of a website. We filter the internal resources from the set of \textit{all} resources by matching the TLD of the website to that of the resource, checking the presence of the TLD of the resource in the SAN List of the website, and comparing the SOA records of the website and the resource, a match in any of the three cases indicates an internal resource. We additionally use public suffix lists~\cite{suffixlist,webxray} to identify remaining internal resources if there are any. 

To find third-party dependence we find the CNAMEs of all the internal resources of a website and obtain the set of CDNs used by the internal resources from a preexisting CDN-CNAME map. The process is summarized in the third condition of Algorithm \ref{alg:code2} (\S\ref{sec:appendix}). Lastly, we determine whether each CDN that hosts the internal resources of a given website is a private service or third-party provided one. For each \textit{(website, CDN)} pair, we extract the CNAMEs of the internal resources of the website which uses that CDN.  Then for each of these CNAMEs, if the TLD of the CNAME is the same as the TLD of the website, we classify the CDN as private. If the website uses HTTPs and the TLD of the CNAME is present within the SAN list obtained from the SSL certificate of the website, the CDN is again classified as private. We finally label the remaining websites by comparing the DNS SOA records of both the website and the CNAME; a mismatch here indicates a third-party CDN. So to classify whether each CDN used by the internal resources of the website is private or third-party service, we follow the same three step heuristics of Algorithm \ref{alg:code1} (\S\ref{sec:appendix}) using the CNAMEs of the internal resources as the \textit{service.url}.

\paragraph{CDN Redundancy} As with DNS measurements, we measure websites that are $(i)$ redundantly provisioned by CDNs (host content from more than one CDN), $(ii)$ critically dependent on a third-party CDN (host content from that one CDN), $(iii)$ use multiple third-party CDNs and $(iv)$ use both private and third-party CDNs. 
 
\subsubsection{CA Measurements.}

For each website that supports HTTPS, we want to find its CA and also identify whether the CA is a third-party or private CA. In addition, we want to know if the website has enabled OCSP stapling, which means that clients do not need to explicitly contact the CA before accessing a site because the certificate's revocation status comes included with the TLS/SSL handshake. This reduces the criticality of the third-party dependency which means an outage of OCSP responders and CDPs does not translate into the website becoming inaccessible. 
To this end, we first make a request using OpenSSL to find a website's listed CA. If the request, which is to port 443, fails, then we assume the website is HTTP-only. At this stage, if the request succeeds, we also check if it has enabled OCSP stapling through information in the request response. The second condition of Algorithm \ref{alg:code2} summarizes our heuristic when the service type is CA.

Next, we find the CA's url from the name of the CA. To classify the CA's url as third-party or not, we make use of the same three step heuristics described in Algorithm \ref{alg:code1} in order to prevent misclassification by using a single approach. If the TLD of the website and the CA's url match, then we classify the CA as private. If there is a mismatch, but if the TLD of CA's url is in the SAN list for the website, then we also classify the CA as private. Finally, if neither of the previous two conditions are met, we check if the DNS SOA record for the CA and the website match. If they do not match, then we classify the CA as third-party. If a website does not fit within any of the previous conditions, then we classify the CA as unknown. 

\subsubsection{Third-party Service Centralization.}

We are particularly interested in the degree of service centralization in markets around the world. The hypothesis is that third-party dependencies and centralization are positively correlated (i.e., high degrees of centralizations in markets with high level of third-party dependencies). However, different markets could be centralized around different or the same set of key service providers.

To measure the degree of centralization across each service, we find the number of third-party websites served by the top-1, top-3 and top-5 providers of each service across the countries and the websites critically dependent on these top providers. We analyze market trends across infrastructures and countries in Sec.~\ref{sec:Findings}.

\subsubsection{Indirect Dependencies.}

Apart from direct dependence of websites on third-party infrastructures like DNS, CA and CDN, these services also rely on other third-party services. So even if a website relies on a private DNS, CA or CDN service but if, for instance, a private CDN uses a third-party DNS then the website is also now (indirectly) dependent on a third-party service. 

To measure these transitive dependencies, like Kashaf et. al \cite{kashaf2020analyzing}, we look at $(i)$ CDN$\rightarrow$DNS dependence, $(ii)$ CA$\rightarrow$CDN dependence and $(iii)$ CA$\rightarrow$DNS dependence across our countries for the top-500 regional sites. 

To measure CDN$\rightarrow$DNS dependence, we collect a list of CDNs used by the top-500 regional sites in each country and find their CNAMEs using the CDN-CNAME map. Then we find the nameserver of each cname and to classify the CDN as private or third-party, we use the heuristics described in Algorithm \ref{alg:code1} (\S\ref{sec:appendix}), where the cname is passed as \textit{w} and each nameserver is passed as \textit{service.url}.

We do the same for CA$\rightarrow$DNS dependence, we collect all CAs used by the top-500 regional sites in each country and find their CA urls. Then we find the nameserver of each CA url and classify the CA as private or third-party. To do this, we again use the heuristics described in Algorithm \ref{alg:code1}, where the CA url is passed as \textit{w} and each nameserver is passed as \textit{service.url}.

Finally, for CA$\rightarrow$CDN dependence, we use the CAs collected and first find the CNAMEs of the CA urls. We then find the CDNs from the CNAMEs using the CDN-CNAME map and classify them as private or third using the technique in (\S\ref{subsec:cdn})

\subsection{Dataset}

The final set of regional websites collected for the 50 countries includes a total of 25,000 websites with 15,774 unique sites. From the 15,774 total unique websites, 15,600 use CDNs and 17,001 uses HTTPs. For our CDN analysis, we found a total of 1,339,871 unique resources and used 877,337 unique internal resources. Across all countries, we found 68 unique third-party CAs, 63 unique third-party CDNs and 764 unique third-party DNS Providers.

%% file: sections/results.tex
\section{Dependency and Centralization}
\label{sec:Findings}

In the following paragraphs, we use the collected dataset to study the degree of third-party dependencies, critical dependency and market centralization around the world. Our analysis looks to answer the following questions:

\begin{itemize}
	\item How common is third-party dependency of websites in different countries and regions around the world?
	\item How much of this dependency is critical, dependent on a single third-party provider, across DNS, CA and CDN infrastructures?
	\item How concentrated is the market of third-party service providers for the different services in our analysis -- DNS, CDN and CA -- within a country, region and globally? 
\end{itemize}

We first look at DNS third-party dependencies, including critical dependencies, around countries and regions.

\subsection{DNS Findings}
\label{subsec:DNS}

\begin{figure}[ht]
	\centering
	\includegraphics[width=\linewidth]{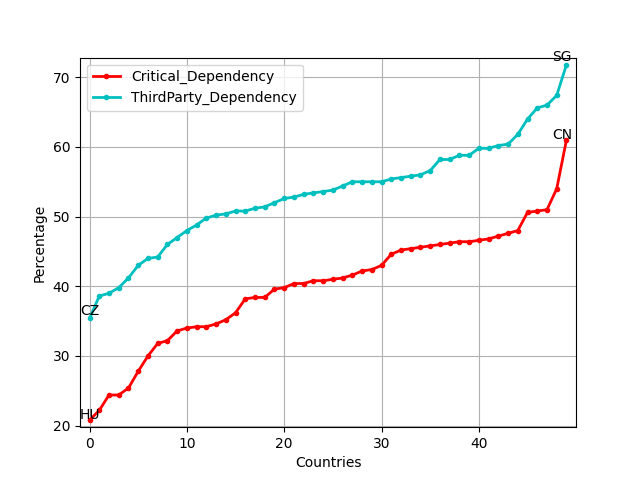}
	\caption{Percentage of a country's top-500 sites using third-party DNS provider and critically dependent on third-party DNS provider. }
	\label{fig:3rdPartyDNS}
\end{figure}

Figure~\ref{fig:3rdPartyDNS} plots the degree of dependency on third-party DNS services for the top-500 
regional sites across countries. We find an average third-party dependency of 53.3\%, and most noticeable 
a wide range of dependency, from as low as 35.4\% for the Czech Republic, to as high as 71.8\% for 
Singapore. 

The figure also shows the percentage of websites, for the different countries, that are critically dependent 
on a single third-party DNS service. While critical dependency is significantly lower than third-party 
dependency, with an average of 40.1\%, the spread is similar ranging from Hungary lowest critical dependency 
of $\approx$21\% to the critical dependency of China close to 61.0\%.

\begin{figure}[ht]
	\centering
\includegraphics[width=\linewidth]{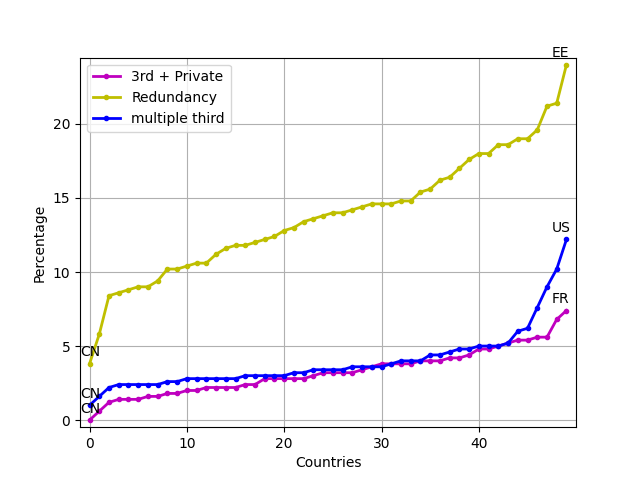}
\caption{Percentage of a country's top-500 sites with redundant DNS, multiple third-party DNS, and both third-party and private providers.}
\label{fig:DNSPlots}
\end{figure}

We also characterize the fraction of websites that use multiple DNS providers (redundant), that use multiple 
third-party DNS providers, and that use both third-party and private DNS providers. Figure~\ref{fig:DNSPlots} plots
these percentages across the top-500 websites of the different countries.



We see that, on average, 14\% of Alexa regional sites have redundant DNS, with Estonia having the maximum 
redundancy of 24\% and China having minimum redundancy of 3.8\%. We find that 4.0\% of sites, on average, 
have multiple third-party DNS providers with the US having a maximum of 12.2\% and China having a minimum 
of 1.0\%. On average, 3.3\% sites use third-party and a private DNS service with France having a maximum value 
of 7.4\% and China having a minimum value of 0.0\%.

\begin{figure}[ht]
    \centering
    \includegraphics[width=\linewidth]{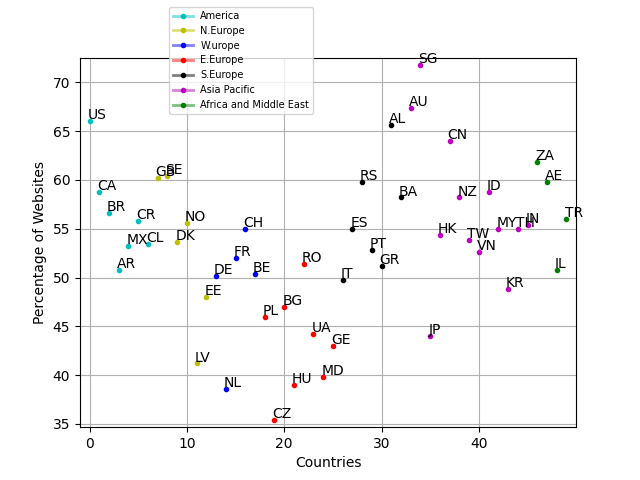}
    \caption{Percentage of each country's top 500 sites that use at least one third-party DNS provider with countries grouped based on regions. Some countries of Europe have significantly lower DNS third-party dependency.}
    \label{fig:regionDNS}
\end{figure}

Next, we cluster the countries in different regions and plot the percentage of websites using third-party DNS in each country to visualize the difference in third-party reliance across locales. Figure~\ref{fig:regionDNS} shows that Eastern Europe has lower DNS third-party dependency.

Finally, we identify the level of dependency of websites on three most popular DNS providers across countries. Table~\ref{tab:popular_DNS} shows the average number of websites relying on each DNS provider. We find that only three DNS providers are used by an average of 76.5\% of the websites that use third-party DNS across countries. Taiwan has the most DNS Centralization with the top three DNS providers being used by 87.7\% of the country's websites that use third-party DNS providers and Hungary has the least DNS centralization (57.9\%). Table~\ref{tab:regional_popular_DNS} shows top-3 popular DNS providers across all regions of our vantage points and the average number of websites dependent on them. We see that in Europe, Asia Pacific and Africa and Middle East, Cloudflare is the most popular third-party DNS provider, whereas in America Amazon Route 53 is the most popular provider.

We compare our findings for the US top-100 and top-500 sites with Kashaf et al.'s \cite{kashaf2020analyzing} a year later. We observe an increase in third-party dependency and redundantly provisioned websites and no change in critical dependency and the percentage of third and private DNS providers. The details of this analysis is included in (\S\ref{sec:appendix}).


\begin{table}[ht]
\centering\small
\caption{Top three DNS providers across regions with the market of DNS providers varying across regions.} 
\label{tab:regional_popular_DNS}
\begin{tabular}[tp]{ |l|c|c|c| }
\hline
Region & DNS Providers &Avg \% of websites & Std Dev\\
\hline \hline
America & \makecell[l]{Amazon Route 53 \\ Cloudflare \\ Akamai} & 77.8 & 83.9 \\
\hline
N.Europe & \makecell[l]{Cloudflare \\ Amazon Route 53 \\ Akamai} & 74.5 & 66.0 \\
\hline
W.Europe & \makecell[l]{Cloudflare \\ Amazon Route 53 \\ Akamai} & 69.1 & 44.2 \\
\hline
E.Europe & \makecell[l]{Cloudflare \\ Amazon Route 53 \\ NsOne} & 72.1 & 65.1 \\
\hline
S.Europe & \makecell[l]{Cloudflare \\ Amazon Route 53 \\ NsOne} & 77.1 & 87.4 \\
\hline
Asia Pacific  & \makecell[l]{Cloudflare \\ Amazon Route 53 \\ Akamai} & 81.2 & 124.7 \\
\hline
\makecell[l]{Africa and \\ Middle East} & \makecell[l]{Cloudflare \\ Amazon Route 53 \\ Akamai} & 78.8 & 61.3 \\
\hline
\end{tabular}
\end{table}

\begin{table}[ht]
\centering\small
\caption{Five most popular third-party DNS providers across the countries with Cloudflare and Amazon Route 53 dominating the market.} 
\label{tab:popular_DNS}
\begin{tabular}[tp]{ |l|c|c| }
\hline
DNS Provider & \makecell[l]{Avg \% of \\ websites} & Std Dev\\
\hline \hline
CloudFlare & 43.4 & 13.9 \\
\hline
\makecell[l]{Amazon \\Route 53} & 30.0 & 10.4 \\
\hline
NsOne & 8.4 & 3.1 \\
\hline
Akamai & 7.6 & 5.0\\
\hline
Dyn & 5.1 & 2.0\\
\hline
\end{tabular}
\end{table}
\begin{table}[ht]
\centering\small
\caption{Five most popular third-party CA providers across the countries. Digi Cert significantly dominates the market} 
\label{tab:popular_CA}
\begin{tabular}[tp]{ |l|c|c| }
\hline
CA Provider & \makecell[l]{Avg \% of \\ websites} & Std Dev\\
\hline \hline
Digi Cert & 36.3 & 7.3 \\
\hline
IdenTrust Inc. & 15.9 & 7.9 \\
\hline
\makecell[l]{Comodo \\CA Limited} & 13.7 & 4.2\\
\hline
GlobalSign & 12.9 & 4.5\\
\hline
\makecell[l]{Starfield \\Technologies, Inc.} & 5.8 & 3.2\\
\hline
\end{tabular}
\end{table}

\subsection{CA Findings}

We find that the average percent of sites using HTTPS across countries was 69.0\%. This average is dragged down by countries such as Argentina, which only had 50.2\% of websites using HTTPS. The United States has the highest rate of HTTPS adoption, with 82.8\% of the Top 500 sites using HTTPS.

\begin{figure}[ht]
  \centering
  \includegraphics[width=\linewidth]{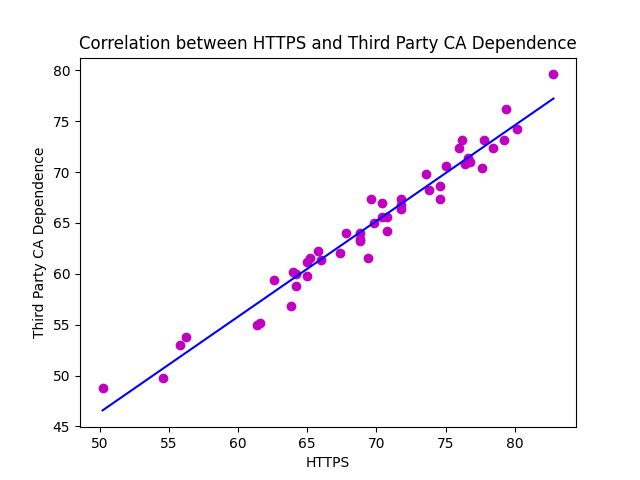}
  \caption{Scatter plot of countries top-500 website using HTTPS (x-axis) and percentage sites using a third-party CA (y-axis).}
  \label{fig:third_vs_https}
\end{figure}

\begin{figure}[ht]
  \centering
\includegraphics[width=\linewidth]{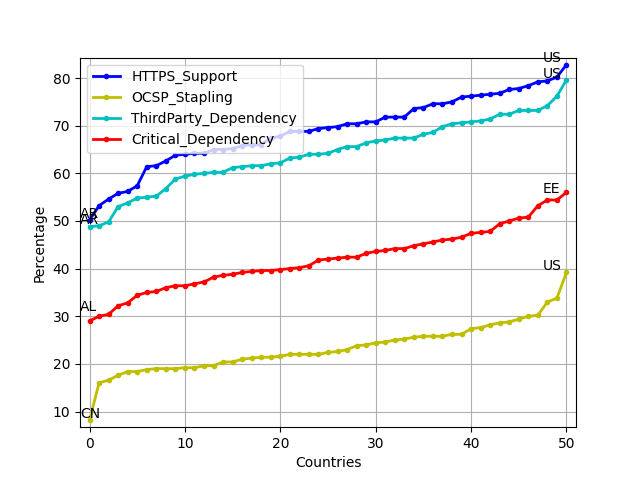}
\caption{Percentage of countries top-500 websites using HTTPs, OCSP stapling, dependent on third-party and critically dependent on a third party CA.}
\label{fig:CAPlots}
\end{figure}

In terms of third-party dependency, 64.3\% percent of sites within our dataset use a third-party CA. The results again ranged from Argentina at the bottom with 48.8\% of its top-500 sites, and the US at the top with 79.6\% of its top 500 sites using a third-party CA (Fig.~\ref{fig:CAPlots}).  Support for HTTPS and use of a third-party CA are strongly positively correlated, as seen in Fig.~\ref{fig:third_vs_https}. On average, nearly all (93.1\%) of HTTPS-supporting websites also use a third-party CA, with standard deviation of only 2\% and no outliers.
 

OCSP stapling is much less popular. On average, 23.3\% of countries' top-500 sites use OCSP stapling with China having the 
lowest usage with only 8.2\% and the US has highest usage at 39.2\%. 

In the context of CA, we define a critically dependent website as a site that $(1)$ uses HTTPS, $(2)$ uses a third-party CA, and $(3)$ does not use OCSP stapling. On average, we find that 42.0\% of the top-500 sites across countries is critically dependent (Fig.~\ref{fig:CAPlots}). The country with the lowest percentage of CA criticality in their top-500 is Albania at 29\%, and the one with the highest is Estonia at 56\%.


\begin{figure}[ht]
    \centering
    \includegraphics[width=\linewidth]{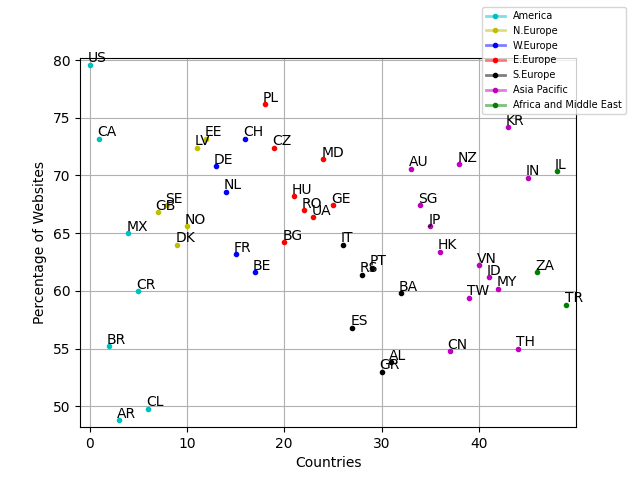}
    \caption{CA third-party dependency across regions.}
    \label{fig:CA_region}
\end{figure}


Figure~\ref{fig:CA_region} plots the countries in different regions and the percentage of websites using third-party CA in each country. 

DigiCert is the most popular CA in all countries' Top 500 Alexa sites except for Estonia, Latvia and Moldova. 36.3\% of websites in our dataset used DigiCert (including Baltimore CyberTrust certificates, which were purchased by DigiCert). The other popular CAs are  IdenTrust Inc. with 15.9\% popularity, Comodo CA Limited with 13.7\% popularity and GlobalSign with 12.9\% popularity. Table~\ref{tab:popular_CA} shows the average number of websites relying on each CA provider. Figure \ref{tab:regional_popular_CA} shows top-3 CA providers across all regions of our vantage points and the average websites using them. We see that GlobalSign is one of the top-3 providers in America, Asia Pacific, Western Europe and in Africa and Middle East but not in Northern Europe, Eastern Europe or Southern Europe.


\begin{table}[ht]
\centering\small
\caption{Top three CA popular providers across regions. Across all regions more than 64\% websites dependent on top-3 CAs. The dominance of top CA providers varies across regions.} 
\label{tab:regional_popular_CA}
\begin{tabular}[tp]{ |l|c|c|c| }
\hline
Region & CA Providers & Avg \% of websites & Std Dev\\
\hline \hline
America & \makecell[l]{DigiCert \\ GlobalSign \\ Comodo} & 67.1 & 192.1 \\
\hline
N.Europe & \makecell[l]{DigiCert \\  IdenTrust Inc. \\ Comodo} & 68.9 & 169.5 \\
\hline
W.Europe & \makecell[l]{DigiCert \\  IdenTrust Inc. \\ GlobalSign} & 64.7 & 124.0 \\
\hline
E.Europe & \makecell[l]{DigiCert \\  IdenTrust Inc. \\ Comodo} & 73.7 & 220.3 \\
\hline
S.Europe & \makecell[l]{DigiCert \\  IdenTrust Inc. \\ Comodo} & 71.0 & 170.3 \\
\hline
Asia Pacific & \makecell[l]{DigiCert \\ GlobalSign \\ Comodo} & 71.6 & 411.7 \\
\hline
\makecell[l]{Africa and \\ Middle East} & \makecell[l]{DigiCert \\ GlobalSign \\ Comodo} & 66.7 & 93.1 \\
\hline
\end{tabular}
\end{table}

Countries, such as Czech Republic, China and Serbia tend to be the most centralized around popular CAs, with more than 80\% of websites using third-party CA choosing one from the top three CAs in their country.  Other countries like Taiwan and Switzerland show less centralization: for these countries, less than 60\% of websites use third-party CA from the top three. 

Finally, in total we identified 68 unique CAs, 15\% higher than the 59 CAs reported in Kashaf et al. for the top 100K sites. We hypothesize this is because we have better representation of websites from different countries and thus better representation of country-specific CAs. Some examples of country-specific CAs that we find include TWCA, a Taiwanese CA, and Microsec Ltc., a Hungarian CA.

We observe that, for top-100 US Alexa sites, HTTPs support and third-party dependency for CAs is the same as Kashaf et al.'s \cite{kashaf2020analyzing} report, even a year later. However, the support for OCSP stapling has increased as shown in (\S\ref{sec:appendix}).

\subsection{CDN Findings}

\begin{figure}[ht]
  \centering
  \includegraphics[width=\linewidth]{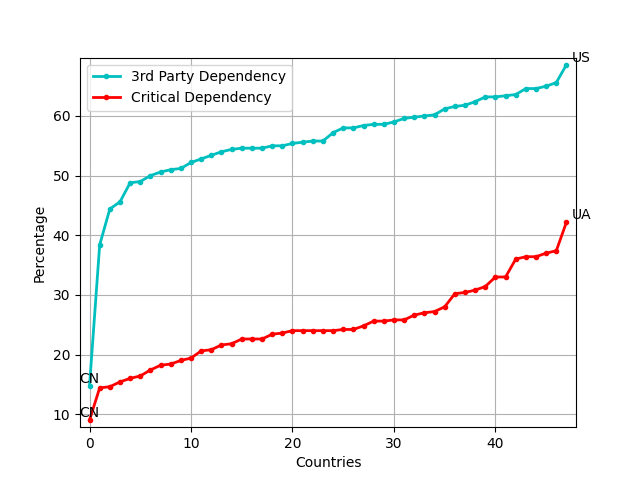}
  \caption{Percentage of websites that rely on third-party CDNs and are critically dependent on a CDN for each country. Difference across countries varies significantly.}
  \label{fig:tp}
\end{figure}

For the first part of our analysis, we show the percentage of websites that are dependent on third-party CDNs (Fig.~\ref{fig:tp}). For each country, we measure using the top 500 websites of that country. Our results show that the average third-party CDN dependency across all countries comes out to be 55.9\%. The country with the lowest dependency of 14.8\% is China while the country with the highest dependency of 68.6\% is the US.

Next, we aim to see the criticality on these CDNs. Critical dependency means when a website is solely hosted on one CDN. Figure~\ref{fig:tp} shows that the country with maximum critical dependency on a third-party CDN is Ukraine with a value of 42.2\% and the country with minimum value of critical dependency is China with a value of 9.0\%. 

\begin{figure}[ht]
  \centering
\includegraphics[width=\linewidth]{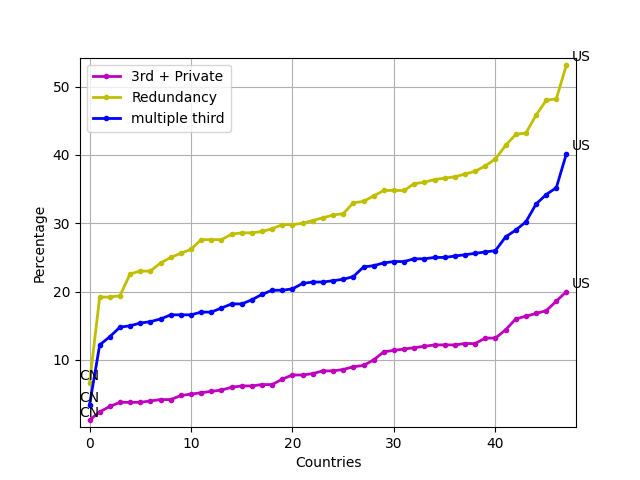}
\caption{Percentage of a country's top-500 sites with redundant CDNs, multiple third-party CDNs, and websites that are hosted on both private and third-party CDNs.}
\label{fig:three}
\end{figure}

We also show the similarity between the CDN usage trends - the percentage of websites using more than one CDN, the percentage of websites using only third-party CDNs, and the percentage of websites using both private and third-party CDNs, as shown in Figure~\ref{fig:three}. On average 32.0\% of Alexa regional sites were redundantly provisioned with CDNs, with the US having the maximum redundancy of 53.2\% and China having minimum redundancy of 6.6\%. 21.9\% sites on average use multiple third-party CDN providers with the US having a maximum of 40.2\% and China having a minimum of 3.4\%. 9.1\% sites on average use third-party and private DNS with the US having a maximum value of 20.0\% and China having a minimum value of 1.2\%.
The Table~\ref{tab:popular_Cdns} shows the top 5 most popular CDNs across the countries we used for measurement, along with the average percentage of websites hosted on them along with the standard deviation. This shows that 72\% of the top 500 Alexa sites use Google as their CDN, 27\% use Amazon Cloudfront and 25\% use Akamai.

\begin{figure}[ht]
  \centering
  \includegraphics[width=\linewidth]{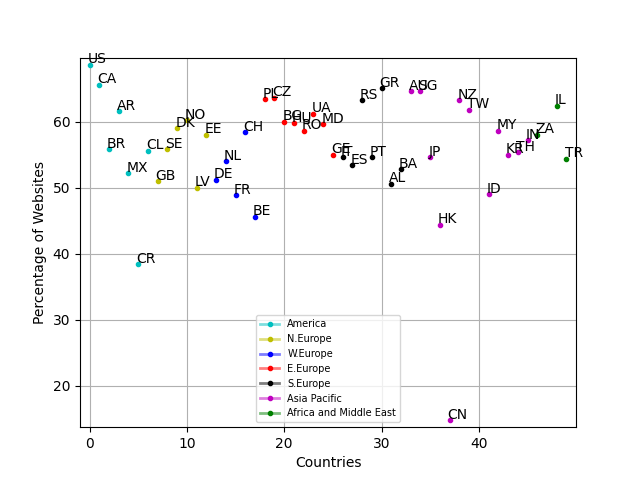}
  \caption{CDNs third-party dependency across regions. Some countries of Europe have significantly lower third-party dependency whereas other European countries have significantly higher dependency. Third-party reliance is country specific.}
\label{fig:regionCDN}
\end{figure}

We find that on average only three CDNs are used to host an average of 91.2\% of the websites that use third-party CDN across countries. South Korea has the most CDN Centralization with the top three CDN providers hosting 95.3\% of the country's websites that are served by a third-party CDN and China has the least centralization with 59.4\% of the country's websites served by third-party CDNs, being hosted on three most popular CDNs. These results demonstrate the high degree of centralization when it comes to CDNs. This along with the fact that most websites rely on third-party CDNs, shine light on the fact that these websites would be highly vulnerable in the face of an attack. However, because the critical dependency is generally low, it would require an attack of several CDNs to shut down any one website. Figure~\ref{fig:regionCDN} plots the percentage of websites using third-party CDNs in each country across different regions. 

Figure~\ref{tab:regional_popular_CA} shows top-3 CDN providers that are popular across different regions of our vantage points and the average percentage of websites using them. The results show that for all regions top-3 CDN providers i.e. Google, Amazon Cloudfront and Akamai are the same and Google is the top provider across all regions. On average, more than 90\% of websites that use a third-party CDN use the top-3 CDN providers across all regions showing a high degree of centralization. 

In the case of CDNs, we observe a decrease in third-party dependency and critical dependency a year later with top-100 US Alexa sites compared to Kashaf et al. \cite{kashaf2020analyzing}. Whereas more websites are now redundantly provisioned and more websites now use multiple third-party CDNs, as shown in (\S\ref{sec:appendix}).




\begin{table}[ht]
\centering\small
\caption{Five most popular CDNs across the countries measured. Google significantly dominates the CDN market.} 
\label{tab:popular_Cdns}
\begin{tabular}[tp]{ |l|c|c| }
\hline
CDN & Avg \% of websites & Std Dev\\
\hline \hline
Google & 72.0 & 10.8 \\
\hline
Amazon Cloudfront  & 27.6 & 9.6\\
\hline
Akamai & 25.2 & 8.3 \\
\hline
Fastly & 12.7 & 5.4\\
\hline
EdgeCast & 6.3 & 3.9\\
\hline
\end{tabular}
\end{table}

\begin{table}[ht]
\centering
\caption{Top three CDN providers across regions. Google, Akamai and Amazon Cloudfront are the top three most popular CDN providers across all regions with Google being the most popular CDN provider across all regions.} 
\label{tab:regional_popular_CDN}
\begin{tabular}[tp]{ |l|c|c|c| }
\hline
Region & CDN Providers & \makecell[l]{Avg \% \\of websites} & Std Dev\\
\hline \hline
America & \makecell[l]{Google, \\ Amazon Cloudfront, \\ Akamai} & 92.1 & 359.7 \\
\hline
N.Europe & \makecell[l]{Google, \\ Amazon Cloudfront, \\ Akamai} & 90.0 & 277.8 \\
\hline
W.Europe & \makecell[l]{Google, \\ Amazon Cloudfront, \\ Akamai} & 89.0 & 215.2 \\
\hline
E.Europe & \makecell[l]{Google, \\ Akamai, \\ Amazon Cloudfront} & 93.9 & 453.8 \\
\hline
S.Europe & \makecell[l]{Google, \\ Amazon Cloudfront, \\ Akamai} & 91.9 & 382.5 \\
\hline
Asia Pacific & \makecell[l]{Google, \\ Akamai, \\ Amazon Cloudfront} & 89.8 & 489.0 \\
\hline
\makecell[l]{Africa and \\ Middle East} & \makecell[l]{Google, \\ Amazon Cloudfront, \\ Akamai} & 92.6 & 174.1 \\
\hline
\end{tabular}
\end{table}

\subsection{Indirect Dependencies}
We also measure percentage of DNS, CA and CDNs that are dependent on another third-party infrastructure.

When measuring transitive dependencies between \newline CDN$->$DNS infrastructure, we find 28 CDNs, on average, across all our vantage points and 30.3\% of them, on average, are dependent on a third-party DNS. With this indirect dependency, 26 more websites are now dependent on third-party DNS.

Additionally, we find a total of 68 CAs across all countries and an average of 26 CAs per country. Measuring CA$->$DNS dependency, shows that on average 17\% of CAs depend on third-party DNS and measuring CA$->$CDN dependency shows 26.3\% of these CAs depend on a third-party CDN.

Given our smaller dataset across countries, the number of additional websites depending on third-party services due to the indirect dependecies of CA on DNS and CDN did not increase significantly.

%% file: sections/TrendAnalysis.tex
\section{Analysing Trends}
\label{sec:trendAnalysis}

Our analysis revealed a wide range of third-party dependencies across the three services we look at: DNS, CA and CDN, varying from 15\% to 80\% of  the top-500 most popular sites across all countries. In this section, we evaluate several factors that may explain part of the variation observed. 
We specifically look at countries' level of $(1)$ economic development, $(2)$ Internet development, their $(3)$ primary official language, and $(4)$ the set of economic trading partners. In each case, we analyse the correlation between third-party dependency of a country and the different factors, fully aware that no factor, \textit{by itself}, could be sufficient to explain the degree of third-party service dependency.

We use Pearson’s correlation coefficients in our analysis. Pearson’s coefficient gives a value between -1 and 1 with values between 0 and 0.3 indicating weak correlation, between 0.3 and 0.7 indicating moderate positive correlation, and above 0.7 indicating a strong positive correlation.  Negative values similarly indicate negative correlations.

\subsubsection{Economic Development}

To understand the effect economic conditions of a country may have on third-party dependency we look at Gross Domestic Product (GDP) Per Capita, and Knowledge Economy Index (KEI) as indicators of the economic development of a country. GDP is a measure of the country’s economic output per person, while KEI is a score of the degree to which the environment of a country is conducive for knowledge to be used effectively for economic development.

The value of correlation coefficient between GDP and third-party dependency in general and across regions is moderate to strong positive. Overall, we find a correlation coefficient of 0.44, but region-specific correlations are mostly strong positive with the exception of Southern and Eastern Europe. The correlation coefficient for America is 0.95, Western Europe is 0.85, Northern Europe is 0.65, Eastern Europe is 0.29, Southern Europe is -0.55, Africa and Middle East is 0.57, and Asia Pacific is 0.60.
 
Similarly, we find moderate to strong correlation with KEI. The average third-party dependency across these infrastructures is 0.39, while the correlation coefficient for America is 0.85, Northern Europe is 0.79, Africa and Middle East is 0.60, and Asia Pacific is 0.45. Western Europe and Eastern Europe and Southern Europe show weak or negative correlations at 0.33 and -0.61, respectively. 

\subsubsection{Internet Development}

Next, we explore the extent to which the advancement of the Internet can explain the variation in third-party dependency across countries. We use Networked Readiness Index (NRI) and ICT Development Index (IDI) as indicators of the development of the Internet in a region. The World Economic Forum's Networked Readiness Index (NRI), also referred to as Technology Readiness, measures the propensity for countries to utilize the opportunities provided by information and communication technology. The International Telecommunications Union's ICT Development Index (IDI), on the other hand, measures the digital divide and ICT performance across countries.

The value of correlation coefficient between NRI and third-party dependency is 0.38. The correlation coefficient for America is 0.94, Northern Europe is 0.84, Western Europe is 0.55, Eastern Europe is 0.33, Southern Europe is -0.33, Africa and Middle East is 0.42, and Asia Pacific is 0.41. 

The value of correlation coefficient between ICT and average third-party dependency is 0.31. The correlation coefficient for America is 0.76, Northern Europe is 0.83, Western Europe is 0.67, Eastern Europe is 0.27, Southern Europe is -0.17, Africa and Middle East is 0.35 and Asia Pacific is 0.34. 

\subsubsection{Language}

\begin{figure}
  \centering
  \includegraphics[width=\linewidth]{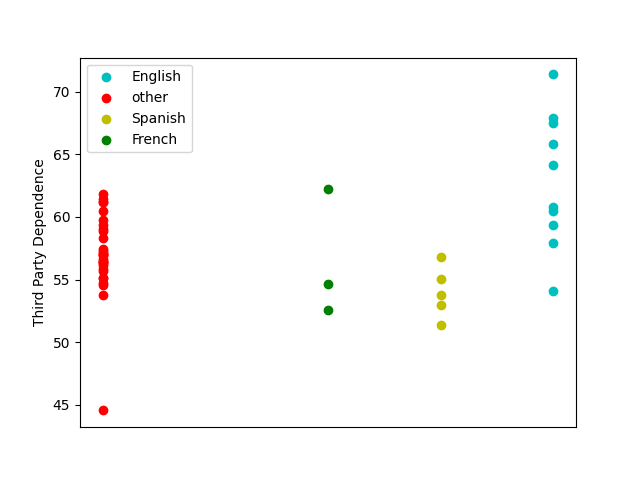}
  \caption{Offical Language (or language that has the de facto status) in each country and the third-party dependency in that country.}
  \label{fig:Lang_3rdPartyDependence}
\end{figure}

To determine the role of language in explaining third-party dependency (e.g., service advertisement may be targeted to specific groups or languages), we use language of the country as an indicator and find the official languages or the language that has the de facto status in each of our vantage points. Then we selected the top 3 languages used by the countries from our dataset and plotted the language used by each country against the third-party dependency in that country. Figure~\ref{fig:Lang_3rdPartyDependence} shows higher concentration of third-party dependency in countries with English as their main language. 



\subsubsection{Trading Blocs}

\begin{figure}
  \centering
\includegraphics[width=\linewidth]{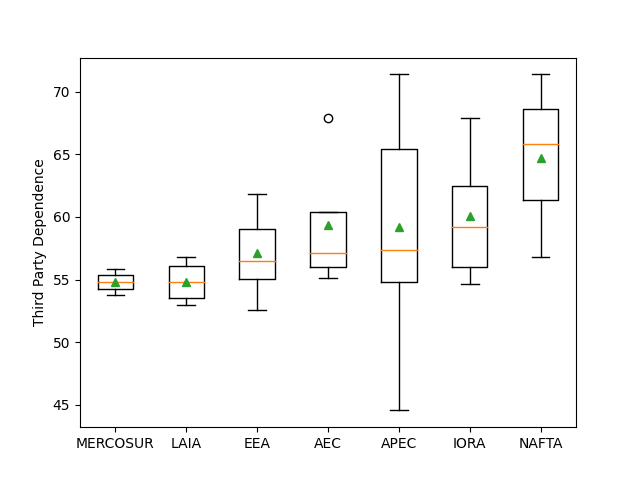}
\caption{Countries in a trading bloc and their percentages of top 500 sites that use a third-party DNS, CA or CDN.}
\label{fig:tradingBoxPlot}
\end{figure}

Last, we explore the association between trading partnerships and third-party dependencies. We hypothesize that countries that share trading partners have similar third-party dependencies as they may rely on common, shared infrastructures. To test this hypothesis, we look at the seven major regional trading blocs in the world economy and then classify the countries based on the trading bloc they belong to, as shown in Table~\ref{tab:tradingblocs_countryCoverage}. These blocs comprise of countries within a specific geographical boundary that have relations to secure regional economic growth. Figure~\ref{fig:tradingBoxPlot} shows the third-party dependency of the countries in each trading block. Countries in the NAFTA block have the maximum dependence and countries in the MERCOSUR and LAIA block have the least third-party dependence. If trading partners were to have a significant role in determining a county third-party dependency, we would expect little variability within the group. This is generally the case but for the APEC block. We note, however, that the set of countries in the APEC trading bloc, including the US, China, Mexico, and Australia, also span several continents, languages and levels of economic development.


\begin{table}[ht]
    \centering
    \caption{Trading Blocs and the countries from our dataset.} 
    \label{tab:tradingblocs_countryCoverage}
    \begin{tabular}[tp]{ |l|l| }
        \hline
        Trading Bloc & Country Codes\\
        \hline \hline
        AEC & \makecell[l]{ID, MY, SG, TH, VN} \\
        \hline
        APEC & \makecell[l]{AU, CA, CL, CN, HK, ID, JP, KR, MX, MY,\\ NZ, SG, TH, US, VN} \\
        \hline
        EEA & \makecell[l]{BE, BG, DE, DK, EE, ES, FR, GR, HU, IT,\\ LV, NL, NO, PL, PT, RO, SE} \\
        \hline
        IORA & \makecell[l]{AE, AU, FR, ID, IN, MY, SG, TH, ZA} \\
        \hline
        LAIA & \makecell[l]{AR, BR, CL, MX} \\
        \hline
        MERCOSUR & \makecell[l]{AR, BR} \\
        \hline
        NAFTA & \makecell[l]{CA, MX, US} \\
        \hline
    \end{tabular}
\end{table}


In conclusion, there is no single factor that explains the differences because each country's advancement in different areas may vary. However, all of these factors partially explain the differences and similarities in third-party dependencies across the countries.

%% file: sections/discussion.tex
\section{Discussion}
\label{sec:Discussion}

Our analysis is subject to a number of limitations. First, Scheitel et al.~\cite{oliver:toplists} raises concern about the ranking opacity and stability of top lists, including the Alexa regional ranking list we rely on for our analysis. While the  Tranco~\cite{pochat:tranco} list addresses some of these issues, we opted for Alexa's regional ranking as this offers the largest regional ranking available to the community, and includes in its listing top regional domain aliases (e.g., yahoo.com.jp and yahoo.com). Tranco allows filter-based selection of regional rankings, however, that list contains an intersection of Tranco's global ranking and domains that also appear in the country-specific Chrome User Experience Report list. This introduces biases in Tranco's regional rankings as Chrome's user Experience list represents only a fraction of users who use Chrome as their browser, have opted in to sync their browsing history, have not set up a Sync passphrase, and have usage statistics enabled~\cite{googleuserexperience}. Between 1\% to 4.5\% of websites appearing in Alexa regional rankings have aliases in different countries (such as google.com and yahoo.com and, for instance,  google.co.rs and yahoo.co.rs in Serbia). These domain aliases can have different dependencies even if they belong to the same entity; for instance rakuten.com uses Akamai as its DNS provider and rakuten.jp uses a private DNS. Alexa's regional rankings includes these domain aliases, while Tranco's regional ranking is derived from its global rankings that may not include these less globally popular, regional versions. We follow recommended best practices~\cite{oliver:toplists}, as Alexa's regional ranking is regarded the best match for our study of global, human-centered study of third-party dependencies, and we make available the downloaded list (including download date) to enable basic replicability. 

Second, we restrict our analysis to top-500 regional sites as Alexa provides rankings for \textit{at least} the top-500 sites per country (i.e., some but not all countries may have more than 500 websites in their rankings). It is possible that the percentage of third-party dependencies for a given country or region change if one could expand the analysis beyond the top-500 regional sites. Exploring the stability of third-party dependencies across subsets of top-regional ranks is part of future work. 

Third, we do not explore underlying physical dependencies and thus our analysis of criticality should be consider a conservative estimate. For instance, websites with redundant DNS services may critically depend on the same underlaying physical infrastructure (e.g., a common landing point in the submarine network). Investigating the relation between third-party dependency, centralization, and critical infrastructure is a promising research direction. 

Last, we limit our analysis of dependency and criticality to 50 out of all possible countries. The set of included countries covers the range in terms of overlap between the top-500 regional sites and the global-ranked lists, has vantage points highly likely located within the claimed country and, together, captures over three-fourth of the  Internet user population. A broader analysis may help better identify the dominant factors behind dependency and centralization; we are exploring 
approaches to this end as part of future work.


%% file: sections/relatedWork.tex
\section{Related Work}
\label{sec:Related Work}

In recent years there has been a move towards Internet centralization and recent work have begun to analyze this in the context of DNS~\cite{zembruzki2020measuring,allman:imc18}, cloud providers~\cite{moura2020clouding}, and the Web in general~\cite{kashaf2020analyzing}. Most recent work has concentrated particularly on DNS centralization. Arkko \cite{arkko:centrality} discusses the implications of the increasing trend towards DNS centralization and encourages that DNS issues be solved without relying on a single DNS service. Livingood~\cite{livingood2019centralized} elaborates on the risks associated with adoption of the latest secure variant of DNS, DNS over HTTPs (DoH), because of its centralized nature and makes recommendations to address that in the DoH implementation and encourages further study prior to widespread adoption. Hoang et al.\cite{hoang2020k} propose a DNS resolution mechanism that shards DNS queries across multiple DoH resolvers to reduce the amount of user’s queries going to a particular DNS resolver and therefore reducing their reliance on any single third party DNS provider.
G. Moura, et al.~\cite{moura2020clouding} measure the degree of DNS centralization by observing the DNS requests coming from the five most popular cloud providers’ DNS resolvers. These cloud providers (CPs) include Microsoft, Google, Facebook, Cloudflare and Amazon. Zembruzki et al.~\cite{zembruzki2020measuring} introduce a methodology and a tool, dnstracker, that allows measurements of the degree of concentration and shared DNS infrastructure by using active DNS measurements. Kashaf et al.~\cite{kashaf2020analyzing} measures the degree of direct and indirect third party dependence of top Global 100k Alexa websites on DNS, CA and CDN infrastructure. 
There have been some attempts to study Internet infrastructures in regions other than North America. Yin et al.~\cite{yin:demystifyingchinacdn} conduct a comprehensive measurement study of China’s CDN infrastructure to understand the development trend of commercial CDNs in this country and their unique characteristics, while Li et al.~\cite{li:mobilehostingchina} use a clustering algorithm to identify hosting providers and infrastructures. Fanou et al.~\cite{fanou:africanweb} look at the availability and utilization of web infrastructure specifically in Africa. Our work builds on Kashaf et al.~\cite{kashaf2020analyzing}, as we leverage the Web to understand third-party dependency and centralization, but aimed instead to explore if and to what extent third-party dependency and centralization varies across countries and regions of the world and what may help explain the observed differences.

%% file: sections/appendix.tex
\appendix
\clearpage
\section{Appendix}
\label{sec:appendix}

\subsection{Countries and Country Codes}

Set of countries included in our analysis, their two-letter code and their percentage of the world's Internet population. 

\begin{table}[h!]
\centering\scriptsize
\label{tab:countries}
\begin{tabular}[tp]{|l|c|r|}
\hline
Country Code & Country & Internet Pop (\%) \\
\hline \hline
AE &United Arab Emirates & 0.19\\
\hline
AL & Albania &0.05\\
\hline
AR & Argentina &0.72\\
\hline
AU & Australia &0.45\\
\hline
BA & Bosnia and Herzegovina &0.05\\
\hline
BE & Belgium &0.22\\
\hline
BG & Bulgaria &0.10\\
\hline
BR & Brazil &3.43\\
\hline
CA & Canada &0.73\\
\hline
CH & Switzerland &0.17\\
\hline
CL & Chile &0.32\\
\hline
CN & China &21.22\\
\hline
CR & Costa Rica&0.08\\
\hline
CZ & Czech Republic & 0.18\\
\hline
DE & Germany & 1.67\\
\hline
DK & Denmark & 0.12\\
\hline
EE & Estonia & 0.02\\
\hline
ES & Spain & 0.91\\
\hline
FR & France & 1.28\\
\hline
GB & United Kingdom & 1.40\\
\hline
GE & Georgia & 0.07\\
\hline
GR & Greece & 0.17\\
\hline
HK & Hong Kong & 0.14\\
\hline
HU & Hungary & 0.16\\
\hline
ID & Indonesia & 4.56\\
\hline
IL & Israel & 0.15\\
\hline
IN & India & 17.89\\
\hline
IT & Italy & 1.08\\
\hline
JP & Japan & 2.50\\
\hline
KR & South Korea & 1.06\\
\hline
LV & Latvia & 0.03\\
\hline
MD & Moldova & 0.07\\
\hline
MX & Mexico & 1.91\\
\hline
MY & Malaysia & 0.54\\
\hline
NL & Netherlands & 0.34\\
\hline
NO & Norway & 0.11\\
\hline
NZ & New Zealand & 0.10\\
\hline
PL & Poland & 0.74\\
\hline
PT & Portugal & 0.16\\
\hline
RO  &Romania & 0.27\\
\hline
RS & Serbia & 0.13\\
\hline
SE & Sweden & 0.21\\
\hline
SG & Singapore & 0.10\\
\hline
TH & Thailand & 0.78\\
\hline
TR & Turkey & 1.33\\
\hline
TW & Taiwan & 0.47\\
\hline
UA & Ukraine & 0.63\\
\hline
US & United States & 6.7\\
\hline
VN & Vietnam & 1.56\\
\hline
ZA & South Africa & 0.68\\
\hline
\end{tabular}
\end{table}

\newpage
\subsection{Heuristics for Third-party Dependency Analysis}

\begin{algorithm}
	\scriptsize
	\caption{ThirdPartyDependence(w)}
	\label{alg:code2}
	\begin{algorithmic}
		\STATE $service ::= DNS | CDN | CA$
		\IF{$\textit{service}=\textit{DNS}$}
		\STATE $NS \leftarrow \textit{digNameservers(w)}$
		\FOR{$ns \in NS$}
		\STATE $nstype \leftarrow \textit{FindserviceType(w,ns)}$
		\IF{$\textit{nstype}=\textit{unknown and concentration(ns) $>$ 50}$}
		\STATE $nstype \leftarrow third$
		\ENDIF
		\ENDFOR
		\ENDIF
		\IF{$\textit{service}=\textit{CA}$}
		\STATE $CA \leftarrow \textit{findCertificate(w)}$
		\STATE $CAURL \leftarrow \textit{findCAURL(w,CA)}$
		\STATE $catype \leftarrow \textit{FindserviceType(w,CAURL)}$
		\ENDIF
		\IF{$\textit{service}=\textit{CDN}$}
		\STATE $IR \leftarrow \textit{findInternalResources(w)}$
		\STATE $cnamesIR \leftarrow \textit{findCnames(IR)}$
		\STATE $CDNs \leftarrow \textit{findCDN(cnamesIR)}$
		\FOR{$cdn \in CDNs$}
		\STATE $cnames \leftarrow \textit{findCnames(w,cdn)}$
		\FOR{$cname \in cnames$}
		\STATE $cnametype \leftarrow \textit{FindserviceType(w,cname)}$
		\ENDFOR
		\ENDFOR
		\ENDIF
	\end{algorithmic}
\end{algorithm}

\begin{algorithm}
	\scriptsize
	\caption{FindserviceType(w,service.url)}
	\label{alg:code1}
	\begin{algorithmic}
		\STATE $service ::= DNS | CDN | CA$
		\STATE $service.type \leftarrow unknown$
		\IF{$\textit{tld(w)}=\textit{tld(service.url)}$}
		\STATE $service.type \leftarrow private$
		\ELSIF{$\textit{isHTTPS(w) and tld(service.url) in SANList}$}
		\STATE $service.type \leftarrow private$
		\ELSIF{$\textit{SOANSProvider(w)} \not= \textit{SOANSProvider(service.url)}$}
		\STATE $service.type \leftarrow third$
		\ENDIF
		\RETURN ${service.type}$
	\end{algorithmic}
\end{algorithm}

\subsection{US Dependency and Centralization}

We look at dependency and centralization with the top-100 and top-500 US Alexa sites and compare our findings to those Kashaf et al.~\cite{kashaf2020analyzing}, and analyze the trends in dependency a year later. Looking at both top-100 and top-500 makes it easier to compare our findings to those Kashaf et al.~\cite{kashaf2020analyzing} as they use top-100 and top-1000 sites in their analysis.

\begin{figure}[!ht]
	\centering
	\subfigure[DNS\label{fig:DNScomparison}]{
		\includegraphics[width=0.4\textwidth]{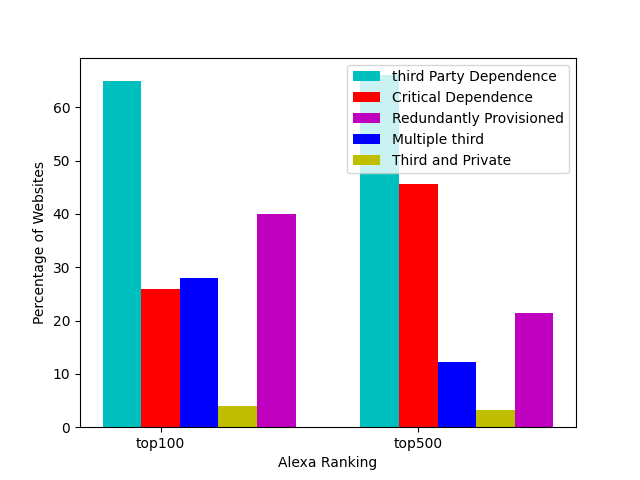}}
	\subfigure[CA\label{fig:CAcomparison}]{
		\includegraphics[width=0.4\textwidth]{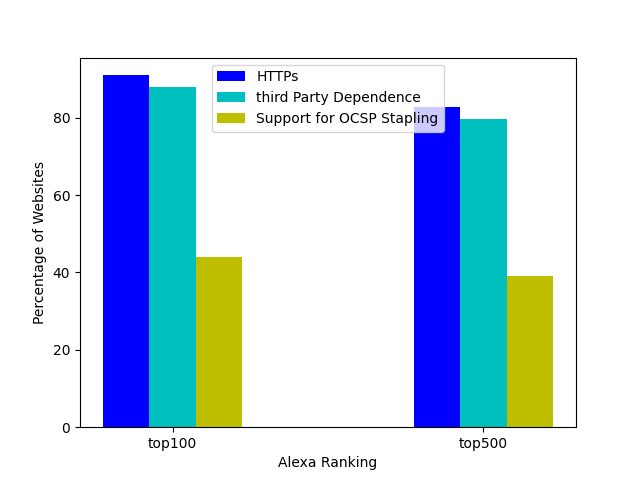}}
	\subfigure[CDN\label{fig:CDNcomparison}]{
		\includegraphics[width=0.4\textwidth]{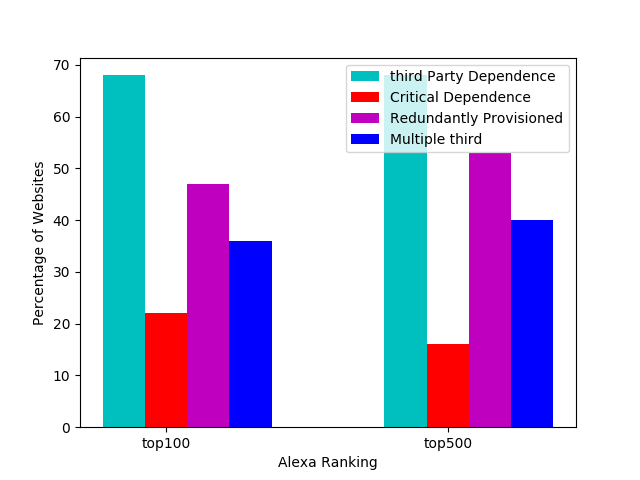}}
	\caption{DNS, CA and CDN metrics for top-100 and  top-500 sites in the US}
\end{figure}

Figure~\ref{fig:DNScomparison} is a bar graph of dependency and redundancy in DNS for the top-100 and top-500 websites. 
We observe an increase in third-party dependency from 48\% to 65\% in the top-100 sites. However, the percentage of websites 
being redundantly provisioned by DNS providers has increased from 21\% to 40\%. There is no change in critical dependence 
and the percentage of third and private DNS providers. We see similar trends between top-100 to top-500 as Kashaf et 
al.~\cite{kashaf2020analyzing} shows from top-100 to top-1000 Alexa sites, i.e. critical dependence increases (26\% to 45.6\%), whereas we see a decrease in websites redundantly provisioned(40\% to 21.4\%) 
and served by multiple third-party providers(28\% to 12.2\%). However, third-party dependency(65\% to 66\%) stays approximately the same.


Figure~\ref{fig:CAcomparison} shows the degree of HTTPs support, third-party dependency and support for OCSP stapling 
for the top-100 and top-500 sites in the US. In this case we observe that our measurements for HTTPs support (91\%) and 
third-party dependency (88\%) match that of Kashaf et al.'s \cite{kashaf2020analyzing} paper for Alexa top-100 sites, however 
the support for OCSP stapling has increased a year later (21\% to 44\%). This indicates a year later, there is less critical 
dependence of CAs on third-party providers. 

Lastly, Fig.~\ref{fig:CDNcomparison} shows third-party and critical dependencies, and the fraction of those redundantly provisioned 
and using multiple third-party CDNs among the top-100 and top-500 websites. For the top-100 websites, we observe a decrease in third-party dependency (76\% 
to 68\%) and critical dependence (43\% to 22\%) a year later. Whereas more websites are now redundantly provisioned by CDNs (32\% 
to 47\%) and more use multiple third-party CDNs(32\% to 36\%).

